\documentclass[%
amsmath,amssymb,
aps,superscriptaddress,pra,notitlepage,reprint
]{revtex4-1}

\usepackage{amsmath,epsfig,color,amssymb}
\usepackage{graphicx}
\usepackage{dcolumn}
\usepackage{bm}
\usepackage{amsmath, amsfonts, amssymb,mathrsfs}
\usepackage{pstricks}
\usepackage{amsxtra}
\usepackage{amsthm}
\usepackage{natbib}
\usepackage{hyperref,color}
\definecolor{darkblue}{rgb}{0.0,0.0,0.7}
\hypersetup{colorlinks,breaklinks,linkcolor=darkblue,urlcolor=darkblue,anchorcolor=darkblue,citecolor=darkblue}
\usepackage{physics}
\usepackage{soul}

\providecommand{\innerp}[2]{\left\langle#1\left\lvert
\vphantom{#1#2}\right.\!#2\right\rangle}  
\providecommand{\comm}[2]{\left[ #1, #2 \right]}  		
\providecommand{\pbrac}[2]{\left\{#1, #2\right\}} 	 	 
\providecommand{\bp}{{\bm{p}}}

\providecommand{\br}{{\bm{r}}}
\providecommand{\bk}{{\bm{k}}}
\providecommand{\bq}{{\bm{q}}}

\begin{document}
\title{Photon thermalization via laser cooling of atoms}
\author{Chiao-Hsuan Wang}
\affiliation{Joint Center for Quantum Information and Computer Science, NIST/University of Maryland, College Park, Maryland 20742, USA}
\affiliation{Joint Quantum Institute, NIST/University of Maryland, College Park, Maryland 20742, USA}

\author{M. J. Gullans}
\affiliation{Joint Center for Quantum Information and Computer Science, NIST/University of Maryland, College Park, Maryland 20742, USA}
\affiliation{Joint Quantum Institute, NIST/University of Maryland, College Park, Maryland 20742, USA}
\affiliation{Department of Physics, Princeton University, Princeton, New Jersey 08544, USA}

\author{J. V. Porto}
\affiliation{Joint Quantum Institute, NIST/University of Maryland, College Park, Maryland 20742, USA}

\author{William D. Phillips}
\affiliation{Joint Quantum Institute, NIST/University of Maryland, College Park, Maryland 20742, USA}

\author{Jacob M. Taylor}
\affiliation{Joint Center for Quantum Information and Computer Science, NIST/University of Maryland, College Park, Maryland 20742, USA}
\affiliation{Joint Quantum Institute, NIST/University of Maryland, College Park, Maryland 20742, USA}
\affiliation{Research Center for Advanced Science and Technology,
University of Tokyo, Meguro-ku, Tokyo 153-8904, Japan}

\begin{abstract}
Laser cooling of atomic motion enables a wide variety of technological and scientific explorations using cold atoms. Here we focus on the effect of laser cooling on the photons instead of on the atoms. Specifically, we show that noninteracting photons can thermalize with the atoms to a grand canonical ensemble with a nonzero chemical potential.
This thermalization is accomplished via scattering of light between different optical modes, mediated by the laser-cooling process. 
While optically thin modes lead to traditional laser cooling of the atoms, the dynamics of multiple scattering in optically thick modes has been more challenging to describe. We find that in an appropriate set of limits, multiple scattering leads to thermalization of the light with the atomic motion in a manner that approximately conserves total photon number between the laser beams and optically thick modes.
In this regime, the subsystem corresponding to the thermalized modes is describable by a grand canonical ensemble with a chemical potential nearly equal to the energy of a single laser photon.
We consider realization of this regime using two-level atoms in Doppler cooling, and find physically realistic conditions for rare-earth atoms.  
With the addition of photon-photon interactions, this system could provide a platform for exploring many-body physics.
\end{abstract}
\maketitle

\section{Introduction}
The laser cooling and trapping of atoms~\cite{Wineland1979,Dalibard1985,Chu1998,*Cohen-Tannoudji1998,*Phillips1998,Metcalf2003} provides a variety of powerful tools for exploring the physics of light and matter~\cite{Bloch2008,Saffman2010,Weimer2010}. While many discussions focus on the atomic behavior, including the thermalization of the motion of a single atom without collisions~\cite{Wineland1979}, curious possibilities regarding the light have also emerged~\cite{Baumann2010}. Simple questions, such as the description of scattered light in optically thick atomic clouds, remain incompletely explored. Another key question is how modification of the photon density of states can change the scattering process. For example, this enables novel regimes of laser cooling in cavities
~\cite{Heinzen1987,Vuletic2000,Vuletic2001a,Curtis2001,Beige2005,Lev2008,Seletskiy2010,Xu2016,Hosseini2017},
and, in interacting systems,
the observation of Bose-Einstein condensation (BEC) of light in semiconductors
\cite{Kasprzak2006,Balili2007,Deng2010,Fleischhauer2008} and molecular dyes
\cite{Klaers2010a,Klaers2010b,Klaers2011,Weitz2013,Kirton2013,Klaers2014,Marelic2015,Schmitt2016,Greveling2017}.
In those cases, the strength of incoherent pumping of excitations determines the photon number and sets a nonzero chemical potential for light~\cite{Klaers2012,Yukalov2012}.
By contrast, in traditional laser cooling, we have a coherent, periodic drive oscillating at the laser frequency. This scenario has been suggested in a general setting as a possible regime of thermalization in a driven system ~\cite{Gullans2016,Wang2018,Seetharam2015} -- leading to a controllable chemical potential for light~\cite{Hafezi2015}. This leads to the natural question of whether similar phenomena can occur in optically thick, laser-cooled atomic systems, where multiple scattering, or cavity confinement, allows laser photons emitted from the atoms to continue to interact with the atomic cloud and potentially thermalize.

Here we partially answer this question by exploring the thermodynamic properties of the photons emitted in the laser cooling process in samples with at least one optically thick axis, comprising many modes.
We show that in this driven-dissipative system the thermalization of these photons arises directly from atomic laser cooling and they are described by a detailed balance condition corresponding to a grand canonical ensemble. These results apply even though the photons are noninteracting, the atoms are noninteracting, and neither is in thermal equilibrium with an external bath. As an illustrative example, this approach allows thermalization of cavity photons with a single atom trapped in an optical cavity. The thermodynamic arguments presented in this work, which are based on the microscopic theory of atom-light interactions, do not rely on specific assumptions about the interaction Hamiltonian or the photonic kinetic energy and, thus, apply to a broad class of many-body photonic systems that can be realized with ultracold atoms.

\begin{figure}[htbp]
\begin{center}
\includegraphics[width=0.4 \textwidth]{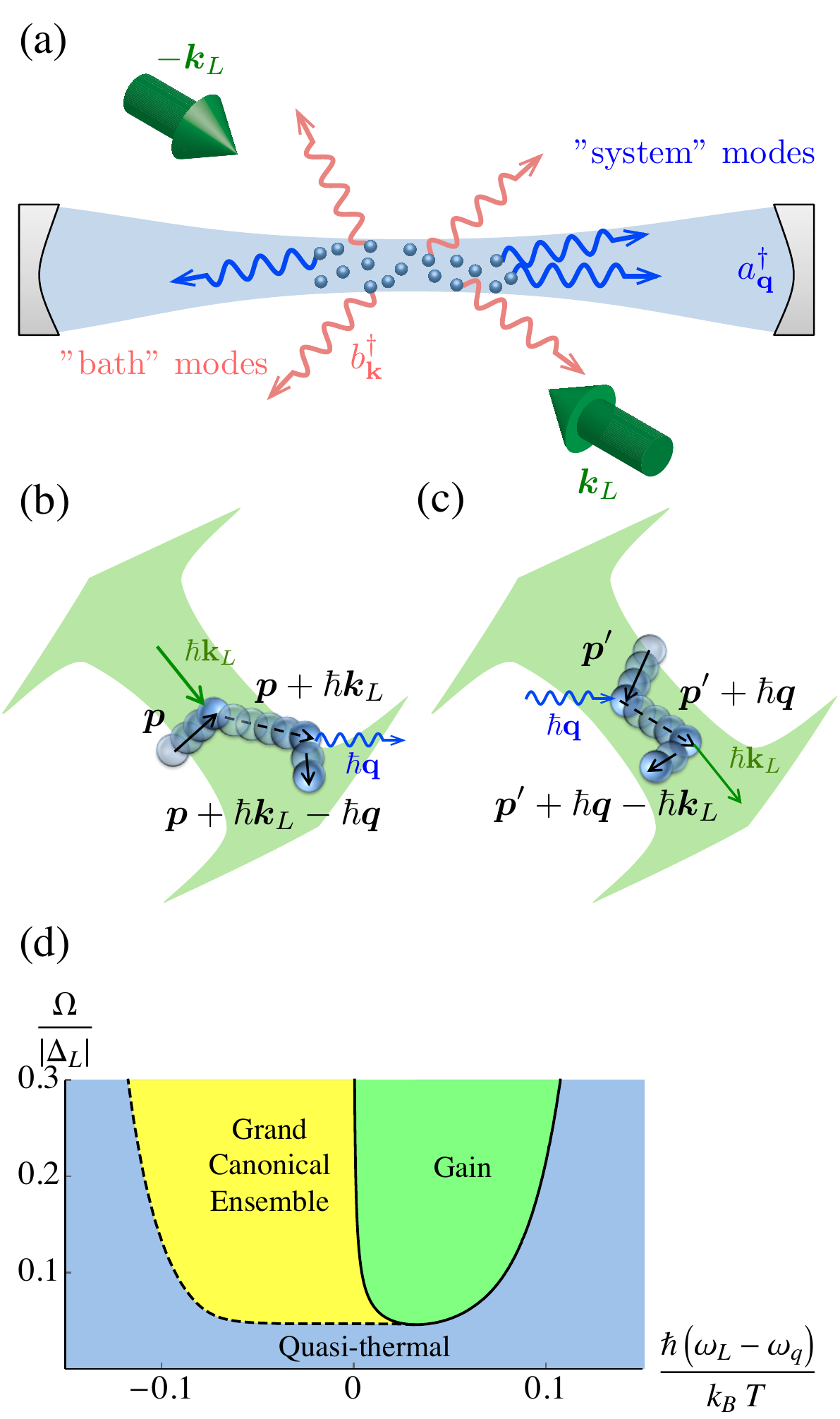}
\caption{(a) Schematic of an ensemble of Doppler-cooled two-level atoms interacting with long-lived cavity (system) photon modes $a_{\bq}$ [blue (dark-gray) wavy arrows within the light-blue (light-gray) region] and lossy (bath) photon modes $b_\bk$ [red (gray) wavy arrows]. Traditional laser cooling arises via loss into the bath modes, while scattering into and out of the blue modes leads to our projected regime of photon thermalization.
(b),(c) The dominant atom-photon scattering processes that lead to a grand canonical ensemble of system photons. (b) The atom is excited by the laser field then emits a system photon. (c) The atom absorbs a system photon then scatters back into the laser field. Effective system plus pump photon number conservation applies due to adiabatic elimination of the atomic excited state and the rotating wave approximation.
(d) Characterization of system photon regimes for a single mode cavity with laser Rabi frequency ($\Omega$) and the laser detuning from the system photon energy ($\hbar(\omega_L-\omega_{\bq})$) as parameters. At higher powers, photon generation can exceed loss as per Eq.~(\ref{GrandL}), leading to either gain [green (gray)] and possibly lasing, or the formation of a grand canonical ensemble for light [yellow (light gray)]. For low powers or large laser detunings from the system photon, photon loss prevents detailed balance with the atomic motion and only quasithermal light is expected [blue (dark gray)].
In this diagram we use the physical parameters for the Yb $~^1 S_0 - ~^3 P_1$ narrow cooling transition~\cite{Ludlow2015} with $\omega_A/2\pi=539$ THz, $\Gamma/2\pi=$ 180 kHz, $\bar{\Delta}_L \approx - 157\, \Gamma$, and assume $\abs{\bk_L-\bq}=\sqrt{2}k_L$.
\label{fig:one} }
\end{center}
\end{figure}

The laser-cooling configuration we focus on in this paper is illustrated in Fig.~\ref{fig:one}(a).  We consider two-level atoms interacting with Doppler-cooling laser beams and two sets of photon modes. One set represents a macroscopic collection of lossy (optically thin), free-space modes and is associated with modes that allow the atom to Doppler cool; we call these ``bath" modes. The photons in the other set of (optically thick) modes are distinguished by the high probability that they will be re-absorbed by the atomic cloud before being lost, either due to intrinsic optical depth (OD) or the existence of a cavity [see Fig.~\ref{fig:one}(a)]. As described above we find that these high OD modes have intriguing thermodynamic properties, and we call them ``system" modes in what follows.

To study the emission and absorption of the system modes during Doppler cooling, we use the quantum jump formalism~\cite{Dalibard1992,Carmichael1993}, but modify it to achieve self-consistent rates with effective elimination of the bath modes.
This allows us to treat the Doppler-cooled atoms as a thermal bath. We then show that the detailed balance condition for photon emission and absorption of the system modes leads to a grand canonical ensemble description of photons at equilibrium, with a chemical potential nearly equal to the energy of a single laser photon. 
We conclude by examining rare-earth atoms as a practical two-level system that can laser cool even at high power.
 We suggest that the rare-earth atoms provide a good platform for realizing thermalization of light using this approach. 
 
The structure of the paper is as follows:  Section~\ref{sec:overview} gives an overview of how photons thermalize in an optically thick laser-cooled atomic ensemble using simple thermodynamic arguments, and contrast this thermalization mechanism with prior work. Section~\ref{sec:twomodecooling} lays out a detailed theoretical formulation of laser cooling with two sets of modes.  Section~\ref{sec:SCFGR} presents a self-consistent analysis of the steady state distribution of system photons, carefully treating the finite lifetime of the atoms as well as possible photon loss mechanisms. Section~\ref{sec:GCE} characterizes the photon steady state  by examining rare-earth atoms as practical two-level atoms to realize the grand canonical ensemble of photons.  Section~\ref{sec:outlook} concludes by motivating the potential theoretical and experimental extensions of our results, including Bose condensation of photons and interacting photonic systems with ultracold atoms.

\section{Overview of theoretical analysis}\label{sec:overview}

In the conventional theory of laser cooling the electromagnetic field is treated as a Markovian bath, which neglects the back-action of the laser-cooling process on the photonic environment. This approximation becomes unjustified when the emitted light from an atom has a high probability of being rescattered by another atom. Such effects are known to play an important role in laser cooling of high optical depth atomic ensembles and are a key limitation in efforts to directly laser cool atoms to quantum degeneracy \cite{Stellmer2013}. 
This regime 
is theoretically challenging because one has to solve 
self-consistently for 
the evolution of the atoms and the rescattered photons.
However, the corresponding interplay between the atom and photon dynamics is central to their thermalization.

To capture this essential physics
we work in the low-excitation limit, such that the nominal Rabi frequency of the cooling laser (2$\Omega$) can be treated perturbatively in $\Omega/|\Delta_L+i \Gamma/2|$ for a laser detuning $\Delta_L=\omega_L-\omega_A$ to an atomic transition with frequency $\omega_A$ and linewidth $\Gamma$. In this regime, photons from the cooling laser scatter from the atoms at the rate $\Omega^2 \Gamma/(\Delta_L^2+\Gamma^2/4)$.  When the coupling of the system photons to the atoms is much weaker than the overall coupling of the bath photons to the atoms, this represents the dominant dynamical process. This cools the atomic motion via loss of photons emitted into bath modes, leading to a thermal ensemble with a temperature set by the Doppler limit $k_B T = \hbar (\Delta_L^2+\frac{\Gamma^2}{4})/2 |\Delta_L|$.  

As the atoms approach the Doppler limit, there remains the slower dynamics of the system photons. These photons can undergo a variety of scattering processes including absorption of system photons and reemission into either bath modes or the cooling laser mode, as well as absorption of cooling laser photons and re-emission into system modes.
In general, the rate for each of these processes can vary widely depending on the regime of operation, as discussed in Secs.~\ref{sec:twomodecooling} and \ref{sec:SCFGR}. For large detunings, however, (corresponding to the high-temperature limit for the atoms) we can understand the steady-state distribution of the system photons by appealing to thermodynamic arguments based on detailed balance between the laser-cooled atoms and the emitted system photons.

\subsection{Photon thermalization with a nonzero chemical potential}

In our hierarchy of bath and system modes, the rate of the system scattering processes is small compared to the overall bath photon-laser photon scattering rate which leads to Doppler cooling of atoms. For a sufficiently high OD, at large detuning and high power $|\Delta_L| \gg \Omega \gg \Gamma$, the key processes that determine the slow dynamics of the system photons are the absorption of cooling laser photons and reemission into system modes [Fig.~\ref{fig:one}(b)], and vice versa [Fig.~\ref{fig:one}(c)].

For a given system mode with label $ \bq$ and frequency $\omega_{\bq}$, these emission and absorption processes are associated with an energy transfer of $|\hbar \omega_L-\hbar \omega_{\bq}|$ between the atoms and system photons. Furthermore, when these processes dominate over the loss of the system photons (typically into bath modes), 
these photons effectively equilibrate with the atoms. In this limit, the atoms approach a thermal distribution with temperature $T$ due to the laser-cooling process, and we have the detailed balance condition, $(\bar{n}_\bq+1)\Lambda^+_{\bq,L}=\bar{n}_\bq\Lambda^-_{\bq,L}$, which leads to
\begin{align}
\frac{\bar{n}_\bq+1}{\bar{n}_\bq}=\frac{\Lambda^-_{\bq,L}}{\Lambda^+_{\bq,L}}
=e^{\beta \hbar(\omega_{\bq} -\omega_L)},
\label{GrandL}
\end{align}
where $\beta^{-1}=k_B T$, $\bar{n}_\bq$ is the mean photon number in mode $\bq$, $\Lambda^{+}_{\bq,L}$ is the rate of absorption of laser photons and subsequent emission into the system modes, and $\Lambda^{-}_{\bq,L}$ is the rate of absorption of system photons and subsequent emission into the cooling laser mode. These scattering rates $\Lambda^{\pm}_{\bq,L}$ are proportional to the population of the initial momentum states of the atoms and therefore pick up the Boltzmann factor for the atomic temperature.
For $\omega_{\bq} > \omega_L$, we will have $\bar{n}_\bq=\frac{1}{e^{\beta \hbar(\omega_{\bq} -\omega_L)}-1}$, which corresponds to a bosonic grand canonical distribution with the temperature of the atomic motion and an effective chemical potential $\hbar \omega_L$. 
Effectively, the system photons have come to a thermal equilibrium with the atoms, but in a frame rotating with the laser frequency so that the energy of a laser photon plays the role of the chemical potential. This detailed balance argument applies to interacting photons as well~\cite{Hafezi2015}.

A nonzero chemical potential for photons occurs because these dominating processes conserve the total number of system photons plus cooling-laser photons. The system photons are thermalized through number exchange between laser photons and system photons when scattered from ground-state atoms. This implies that the cooling laser acts as a number reservoir for the system photons, while the atoms play the role of the energy reservoir in the grand canonical ensemble. There are modifications to this picture, derived below, arising from effects such as the finite lifetime of the system photons, that lead to perturbative shifts in the effective temperature and chemical potential.
These corrections arise because the underlying system is still a nonequilibrium, mesoscopic one. 
 We emphasize that this picture of a grand canonical ensemble for system photons is distinct from the trivial effect whereby the scattered light reflects the temperature of the atoms~\cite{Westbrook1990}. In this case, the Gaussian spectrum of scattered light reflects the Maxwell-Boltzmann distribution of the laser-cooled atoms, as opposed to being in a Bose distribution, as we find here.

For $\omega_{\bq} < \omega_L$, in contrast to the case above, there is a runaway process and we expect gain or lasing instead of an equilibrium steady state since it is more probable to emit photons into such system modes than absorb photons from the mode. In an optically thick medium, the system photons are diffusive and become trapped for a finite time related to the OD; however, due to runaway processes the steady state may become dominated by saturation effects, which we do not account for in this work. Restricting the system photon states to $\omega_{\bq} > \omega_L$ by a cavity or other means will prevent gain. For simplicity, we focus on the cavity model in the later discussions.

Reaching the regime where we can safely neglect the loss of the system photons, due to scattering into bath modes or other decay mechanisms, requires a careful consideration of those other, lossy, emission and absorption processes that occur during the laser-cooling dynamics.
The above arguments based on detailed balance require energy conservation during the microscopic energy transfer process between atoms and system photons, while the finite lifetime of the atomic ground state due to the Doppler-cooling process potentially violates this condition. To incorporate the mechanisms leading to Doppler cooling, with the mechanisms leading to detailed balance, we develop a theoretical tool called the self-consistent Fermi's golden rule (SC-FGR). Under the framework of SC-FGR, we can treat Doppler cooling of atoms, all emission and absorption processes of system photons, and the loss mechanisms in a self-consistent manner as described in the following sections. We find that the finite lifetime of the dressed atomic ground state due to Doppler cooling, a necessary ingredient for atomic thermalization, 
modifies the simple detailed balance argument presented above. Specifically, at high detuning and high laser power, where the atomic temperature is far from the Doppler limit, we see grand canonical ensemble (GCE) and other behavior, as summarized in the phase diagram of Fig.~\ref{fig:one}(d).

\subsection{Comparison to previous work}
It is helpful to contrast the results of this paper with previous work on photon thermalization with a nonzero chemical potential, which has a long history.  Such work can be broadly classified into two categories that rely (i) on interactions between light and matter where the matter is in thermal equilibrium with an external reservoir, or (ii) multiple photon-photon collisions mediated by matter. The former includes the earliest theoretical proposal of photon BEC in a plasma \cite{ZelDovich1969}, photon thermalization and condensation in a dye-filled microcavity \cite{Klaers2010a, Klaers2010b,Klaers2012,DeLeeuw2013,Kirton2013}, as well as recent proposals in quantum optomechanics \cite{Weitz2013,Fani2016}; the latter includes photon BEC through photon-photon scattering in a nonlinear resonator \cite{Chiao2000} and BEC of exciton polaritons \cite{Kasprzak2006,Balili2007,Deng2010} and stationary-light polaritons \cite{Fleischhauer2008}. Our approach has the most in common with (i), however, it falls outside this category because the bath for the photons (i.e., the atoms) is not in thermal equilibrium with an external reservoir but rather driven to a nonequilibrium steady state with a thermal description. In optically thick atomic media, the dynamics of the system photons, which are generated during the laser-cooling process, must then be treated self-consistently with the equilibration dynamics of the atoms.

A related class of studies is concerned with characterizing the nonequilibrium steady state of driven-dissipative photonic systems~ \cite{Lebreuilly2018,Kilda2017,Sieberer2013,Tauber2014,Maghrebi2016,Sieberer2016}. In many instances, these systems are driven towards an effectively thermal state at long times. However,  statistical mechanical arguments do not guarantee such emergence of one of the standard thermodynamic ensembles, making the results dependent on underlying assumptions about the system. In cases where universal results can be obtained using the renormalization group \cite{Sieberer2013,Tauber2014,Maghrebi2016,Sieberer2016}, the thermal behavior is only guaranteed to apply at long-time and long-wavelength scales. Although the analysis from these studies does not apply to our system, we find a similar conclusion that, under a broad range of conditions, laser cooling in optically thick media acts as an effectively thermal driven-dissipative system. This result is surprising in the context of laser cooling because one might expect that multiple scattering in such driven optically thick atomic media leads to complicated many-body effects and nonthermal steady states \cite{Walker1990,Labeyrie1999,Guerin2017,Saint-Jalm2018}.

\section{Laser Cooling with Optically Thick and Thin Modes} \label{sec:twomodecooling}
Here we study light scattering in dilute, optically thick atomic gases and neglect radiative dipole-dipole interactions between atoms. The Hamiltonian for a two-level atom interacting with a single laser and two sets of photonic modes $H_S$, $H_B$ is
\begin{align} 
& H = H_S +H_B+H_{AS}+H_{AB}+ H_{AL}(t) + H_A,\label{eqn:H} \\
& H_A =\frac{\bp^2}{2m} + \hbar \omega_A \ket{e}\bra{e}, \\
&  H_S =\sum_{\bq} \hbar \omega_{\bq} a_{\bq}^{\dagger} a_{\bq}, \, H_B = \sum_{\bk} \hbar \omega_{\bk} b_\bk^{\dagger} b_\bk,\\
\label{eqn:HAS}
& H_{AS} = -\sum_{\bq} \hbar \alpha_{\bq}a_{\bq} e^{i \bq \cdot \br} \ket{e} \bra{g}+\text{H.c.},\\
& H_{AB} =-\sum_{\bk} \hbar \beta_{\bk} b_\bk e^{i \bk \cdot \br} \ket{e} \bra{g}+ \text{H.c.},\\ \label{eqn:Hal}
& H_{AL}(t) = -\hbar \Omega e^{-i \omega_L t} e^{i \bk_L \cdot \br} \ket{e} \bra{g}+ \text{H.c.},
\end{align}
Here $H_A$ is the Hamiltonian of a two-level atom, and $m$, $\bp$, $\br$, $\omega_A$ are the mass, momentum, position, and the transition frequency of the atom; $H_{S}$ describes long-lived system photon modes of interest associated with bosonic annihilation operators $a_{\bq}$ and energies $\hbar \omega_{\bq}$; $H_{B}$ describes lossy bath modes with bosonic annihilation  operators $b_\bk$ and energies $\hbar \omega_{\bk}$; $H_{AS}$, $H_{AB}$, and $H_{AL}(t)$ represent atom-system photon, atom-bath photon, and atom-laser interactions. 
Throughout the text $\bq$ is the wave vector for system photons and $\bk$ labels bath photons. $2 \alpha_{\bq}$ and $2 \beta_{\bk}$ are the single-photon Rabi frequencies of the system photons and bath photons.   Note that all coupling frequencies $\alpha_{\bq}$, $ \beta_{\bk}$, and $\Omega$ are assumed to be real and are defined as a half of the usual Rabi frequencies to absorb the $1/2$ factor for notational simplicity.  
 
In what follows we separate the system and bath modes by including cavity end mirrors, and assume that the intrinsic optical depth of the atomic cloud is much smaller than $1$, while the effective optical depth after including the cavity is greater than $1$, so that the system photonic modes are now cavity modes. For simplicity, we make a plane-wave approximation for the cavity modes so that Eq.~(\ref{eqn:HAS}) still holds.  In principle, our general concept of thermalizing system photons via laser cooling may also be realized in a cavity-free setting. 

To obtain an effective Hamiltonian and the corresponding master equation describing the evolution of the atom and the system modes $a_{\bq}$, we first integrate out the lossy bath modes $b_{\bk}$ in the weak excitation limit $\Omega \ll |\Delta_L + i \Gamma/2|$. According to Fermi's golden rule and momentum conservation, the spontaneous emission rate from the atomic excited state with momentum $\bp$ due to all bath modes $b_\bk$ is
\begin{align}
\Gamma_b(\bp)= \frac{2 \pi}{\hbar} \sum_\bk \abs{\hbar \beta_\bk}^2 \delta \left(\Delta E_{eg}(\bk,\bp) \right)\approx \Gamma_b(0) \equiv \Gamma_b,
\label{gammaa}
\end{align}
with the energy difference between the initial excited  and final ground  states defined as
\begin{align}
\Delta E_{eg}(\bk,\bp) = \frac{\bp^2}{2m}+\hbar\omega_A-\frac{\abs{\bp-\hbar \bk}^2}{2m}-\hbar\omega_{\bk}.
\label{Ee}
\end{align}
The effect of atomic motion on the total decay rate is negligible, assuming the atomic transition energy $\hbar \omega_A$ is much larger than the Doppler shift and the recoil energy.

Similarly, the total spontaneous emission rate from the atomic excited state with $\bp$ due to system modes $a_{\bq}$ is
\begin{align}
\Gamma_a(\bp)= \frac{2 \pi}{\hbar} \sum_\bq \frac{\abs{\hbar \alpha_{\bq}}^2 \hbar \kappa_{\bq}}{\Delta E_{eg}(\bq,\bp)^2+\hbar^2\kappa_{\bq}^2/4}\approx \Gamma_a(0) \equiv \Gamma_a.
\label{gammab}
\end{align}
Here $\kappa_{\bq}$ is the cavity decay linewidth of the system photons $a_{\bq}$. Note that the overall approach here can also apply to the case without a cavity by replacing the Lorentzian factor in Eq.~(\ref{gammab}) with a Dirac $\delta$ function. We work in the limit $\Gamma_b \gg \Gamma_a$ so that the spontaneous emission rate into bath modes is approximately the atomic natural linewidth of the atom, $\Gamma_b \approx \Gamma$.

\begin{figure}[htbp]
\begin{center}
\includegraphics[width=0.45 \textwidth]{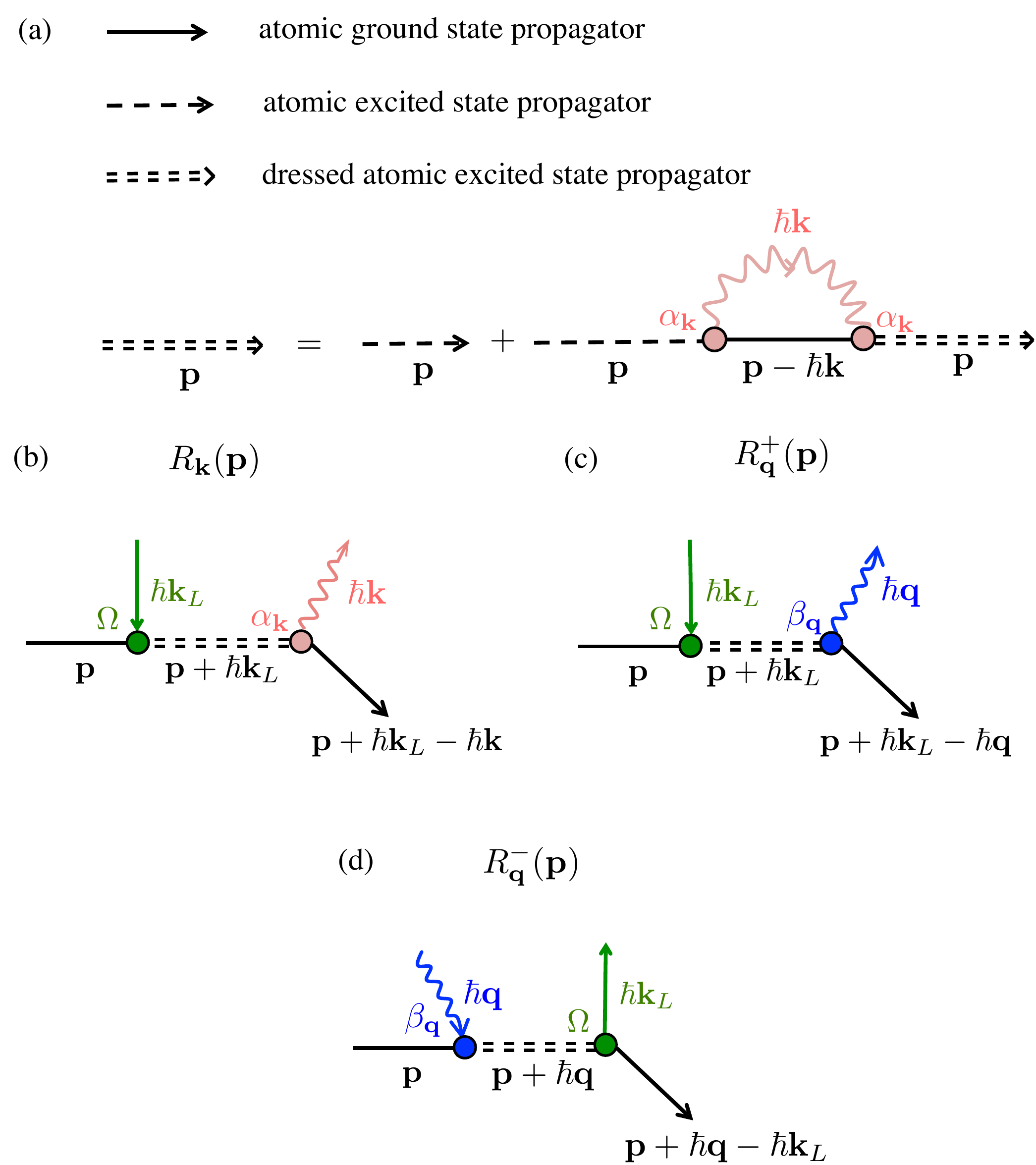}
\caption{Diagrams for laser-induced scattering between atomic ground states with different momenta to lowest order in $\Omega$. (a) The dressed atomic excited-state propagator (double dashed line) is defined by including nonperturbative effects due to the coupling to the bath photon modes $b_\bk$ [red (gray) wavy arrow] and neglecting the effect of the system photons assuming $ \Gamma_b \approx \Gamma \gg \Gamma_a$. (b) The diagrammatic representation of the scattering amplitude of a ground-state atom (solid black line) from an initial momentum state $\bp$ to a final state $\bp+\hbar\bk_L-\hbar\bk$ by absorbing a pump photon [green (gray) straight arrow] and emitting a bath photon. This process is associated with an effective coupling $R_\bk(\bp)$ between momentum states $\ket{g,\bp}$ and $\ket{g,\bp+\hbar\bk_L-\hbar\bk}$. (c) The scattering amplitude from $\ket{g,\bp}$ to $\ket{g,\bp+\hbar\bk_L-\hbar\bq}$ by absorbing a pump photon and emitting a system photon $a_{\bq}$ [blue (dark-gray) wavy arrow], associated with a coupling $R^+_\bq(\bp)$. (d) The scattering amplitude from $\ket{g,\bp}$ to $\ket{g,\bp+\hbar\bq-\hbar\bk_L}$ by absorbing a system photon $a_{\bq}$ and emitting a pump photon, associated with a coupling $R^-_\bq(\bp)$. Not shown is the process in which a system photon is rescattered to a bath photon, which is treated in Fig.~\ref{fig:transitions}.}
\label{fig:feynman}
\end{center}
\end{figure}

Scattering between atomic ground states with different momenta is induced by the laser and the photon modes. Working in the weak excitation limit, we calculate the transition rates with time-dependent perturbation theory to the lowest order in 
$\frac{\Omega}{\Delta_L+i \Gamma/2}$~\cite{Cohen-Tannoudji1992}. The relevant processes are illustrated diagrammatically in Fig.~\ref{fig:feynman}. For example, 
Fig.~\ref{fig:feynman}(b) represents the coupling from an initial ground-state atom in momentum state $\bm{p}$, $\ket{g,\bp}$, to the new momentum state $\bp+\hbar\bk_L -\hbar\bk$, $\ket{g,\bp+\hbar\bk_L-\hbar\bk}$, with an additional emission of a bath photon with momentum $\hbar\bk$ into the $b_\bk$ modes. Using second-order time-dependent perturbation theory, represented diagrammatically in Fig.~\ref{fig:feynman}(b), we get an effective coupling between atomic motional ground states with momentum ${\bp}$ and ${\bp+\hbar\bk_L-\hbar\bk}$ as
\begin{align}
R_{\bk}(\bp)= \frac{\Omega \beta_{\bm{k}}}{  \omega_L-\omega_A -\frac{\bp \cdot \bk_{L}}{m} - \frac{E_r(\bk_L)}{\hbar} +i \frac{\Gamma}{2}}.
\label{Rbk}
\end{align}
Note that the term $-\frac{\bp \cdot  \bk_L}{m}=-\textbf{v}\cdot \bk_L$ is the Doppler shift of the laser frequency as seen by the moving atom. The photon recoil energy, defined as 
\begin{align}
E_{r}(\bk_L)=\frac{\hbar^2 \bk_L^2}{2m},
\label{Er}
\end{align}
also shifts the laser frequency by an additional amount $\frac{\hbar \bk_L^2}{2m}$.

Assuming the magnitude of the atom-bath photon coupling constants $\beta_\bk$ are insensitive to the photon energy over the atomic linewidth, we calculate the total dissipation rate for an atom with momentum $\bp$ due to laser-bath scattering using Fermi's golden rule.  Diagrammatically this is equivalent to summing over the bath output states in Fig.~2(b) labeled by $\hbar \bk$,
\begin{align} 
\gamma^{\rm}(\bp) & = \frac{2 \pi}{\hbar} \sum_\bk \abs{\hbar R_\bk(\bp)}^2 \delta\left(\Delta E_{gg}(\bk,\bp) \right)\notag\\ &\approx  \frac{\Omega^2 \Gamma}{(\bar{\Delta}_L - \frac{\bp \cdot  \bk_L}{m})^2+\frac{\Gamma^2}{4}},
\label{gammap}
\end{align}
with the ground-to-ground energy difference defined as
\begin{align}
\Delta E_{gg}(\bk,\bp)=\frac{\bp^2}{2m} +\hbar \omega_L -\frac{|\bp+\hbar\bk_L-\hbar\bk|^2}{2m} - \hbar\omega_{\bk}.
\label{Eg}
\end{align}
where $\bar{\Delta}_L=\omega_L-\omega_A-E_r(\bk_L)/\hbar$ is the shifted detuning of the laser, including the recoil shift from the bare detuning $\Delta_L$. This momentum-dependent dissipation rate can lead to Doppler cooling of the atomic motions for $\Delta_L < 0$~\cite{Wineland1979,Lett1989}. In Appendix~\ref{sec:mastereq}, we recover the results of the standard Doppler cooling theory applied to our two-mode (system and bath) configuration. 
More generally, when the Doppler-cooled atomic ensemble can be treated as a thermal bath for the system photons, the parametric (laser-induced) coupling between atomic motion and the system photons will bring the system photons to an equilibrium state describable using a grand canonical ensemble, leading to an effective nonzero chemical potential set by the pump frequency $\mu=\hbar \omega_L$~\cite{Hafezi2015}. 

The system photons also give rise to an effective coupling between atomic ground states $\ket{g,\bp}$ and $\ket{g,\bp+\hbar \bk_L-\hbar \bq}$, which
  to lowest order in $\frac{\Omega}{\Delta_L+i \Gamma}$ [see Fig.~\ref{fig:feynman}(c)] is
\begin{align}
R^+_\bq(\bp)=\frac{\alpha_{\bq} \Omega}{\bar{\Delta}_L-\frac{\bp \cdot \bk_L}{m}+i\frac{\Gamma}{2}}.
\label{R+bq}
\end{align}
In contrast to the bath modes, the system modes have high effective optical depth and we must also take into account the reverse process of first absorbing a system photon and reemitting into the laser-cooling field. 
This gives rise to the effective coupling between atomic stateS $\ket{g,\bp}$ and $\ket{g,\bp+\hbar \bq-\hbar \bk_L}$ [Fig.~\ref{fig:feynman}(d)]
\begin{align}
R^-_\bq(\bp)=\frac{\alpha_{\bq} \Omega}{\bar{\Delta}_\bq-\frac{\bp \cdot \bq}{m}+i\frac{\Gamma}{2}},
\label{R-bq}
\end{align}
where $\bar{\Delta}_\bq=\omega_{\bq}-\omega_A-\hbar \bq^2/2m=\Delta_\bq-E_r(\bq)/\hbar$ is the shifted detuning of the system photon, including the recoil shift. The combined effects of these momentum-changing transitions lead to broadening of the motional eigenstates of the atom.

To determine the transition rates leading to the detailed balance condition for the system photons, we require a similar sum over the outgoing states as in Eq.~(\ref{gammap}). 
If we can account for all the relevant processes---including the one not shown in Fig.~\ref{fig:feynman} in which a system photon is rescattered into a bath mode---we would have a complete description of the master equation for the system modes.
However, as we discuss in the next section, this requires a self-consistent treatment of the atomic-ground-state scattering to account for the broadening of the ground-state energies due to the Doppler-cooling process. 

\section{Self-Consistent Calculation of Transition Rates}
\label{sec:SCFGR}
To fully account for the finite lifetime of the motional eigenstates of the atoms in their electronic ground states due to laser-cooling-induced transitions, here we develop a formulation of FGR we call THE \textit{self-consistent Fermi's golden rule} (SC-FGR)~\cite{Pastawski2007,Gumhalter2008}, in which the effect of the rapid dissipation is treated self-consistently. As shown below, this leads to a replacement of the $\delta$ function in the usual sum over atomic states with an energy-broadened approximate $\delta$ function. This allows us to evaluate the rates for system photon emission and absorption, leading to a simple set of rate equations for system modes tracing over the atomic motion. Our SC-FGR approach yields a key result: For experimentally accessible parameters, the atomic temperature must be significantly higher than the Doppler cooling limit for our theory to apply.

\begin{figure}[htbp]
\begin{center}
\includegraphics[width=0.47 \textwidth]{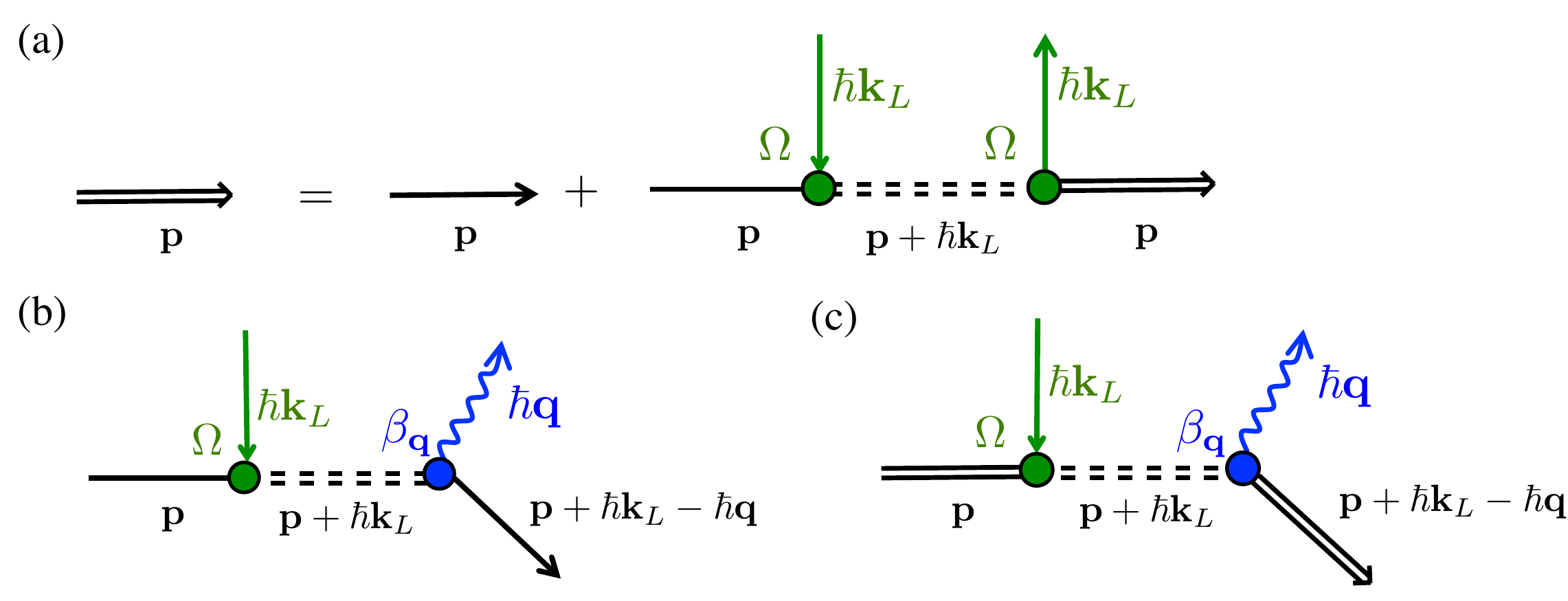}
\caption{Diagrammatic comparison between Fermi's golden rule and self-consistent Fermi's golden rule. (a) The dressed state picture of the atomic-ground-state propagator (double black line) is defined by including nonperturbative effects due to the ground-state scattering induced by laser. The double dashed line is the atomic excited-state propagator dressed by the bath modes $b_\bk$ as shown in Fig.~\ref{fig:feynman}(a). (b) For reference, we give the diagrammatic representation of the (regular) Fermi's golden rule scattering amplitude that involves a system photon $a_{\bq}$ emission, using the dressed excited-state propagators which leads to a standard FGR result. (c) The diagrammatic representation of the self-consistent Fermi's golden rule scattering amplitude, in which we also replace the atomic-ground-state propagators with dressed ones to account for the finite lifetime of the original and final atomic motional states set by rapid emission processes into bath modes.}
\label{fig:SCFGR}
\end{center}
\end{figure}

The general concept of SC-FGR can be understood through an example illustrated in Fig.~\ref{fig:SCFGR}. 
As seen diagrammatically in Fig.~\ref{fig:SCFGR}(a), the ground-state propagator for the atoms becomes dressed with the excited state due to the presence of the Doppler cooling laser field.  Solving this equation self-consistently, we find that the dressed propagator for the ground-state atoms is approximately
\begin{equation} \label{propagator}
 \frac{i}{\omega - \bp^2/2m\hbar + i \gamma(\bp)/2} = \pi \delta_{\gamma(\bp)}(\omega-\bp^2/2m\hbar) + i\text{P.V.},
\end{equation}
where $\delta_{\epsilon}(\omega)=\frac{\epsilon /2 \pi}{\omega^2 + \epsilon^2/4}$ is a broadened $\delta$ function for $\omega $ with width $\epsilon$ and P.V. corresponds to the principal value in the limit $\gamma(\bp)\to 0$.
When the broadening is neglected we recover the usual FGR transition rate shown in Fig.~\ref{fig:SCFGR}(b). In comparison, for our SC-FGR calculation [Fig.~\ref{fig:SCFGR}(c)], we replace the atomic-ground-state propagators with dressed ones. This way, we evaluate the system photon emission process over a finite time before the atoms are reset by the emission process (quantum jump) into bath modes, which leads to a broadening of the $\delta$ function that arises in the standard FGR, as detailed in Appendix B. The treatment of such a lifetime broadening effect is crucial since the detailed balance equilibration to a grand canonical ensemble of photons relies upon energy-conserving transitions between system modes and their (parametrically) coupled bath -- atomic motion, in our case.

In the next two sections, we use the SC-FGR to derive the system photon emission rate, the system photon absorption rate mediated by the laser, and the system photon loss rate due to scattering into bath modes. We find that the system photons follow a grand canonical distribution when the photon losses due to scattering into bath modes or from the cavity mirror are negligible, and the SC-FGR analysis sets an additional high-temperature requirement on the atomic motion: $k_B T \gg \hbar \Gamma/2$. For a reader interested in the microscopic details, in Appendix B we give an alternative derivation of the SC-FGR using the quantum jump picture, which agrees with this diagrammatic analysis.

\subsection{Photon equilibration mediated by dressed atoms}
\begin{figure}[htbp]
\begin{center}
\includegraphics[width=0.47 \textwidth]{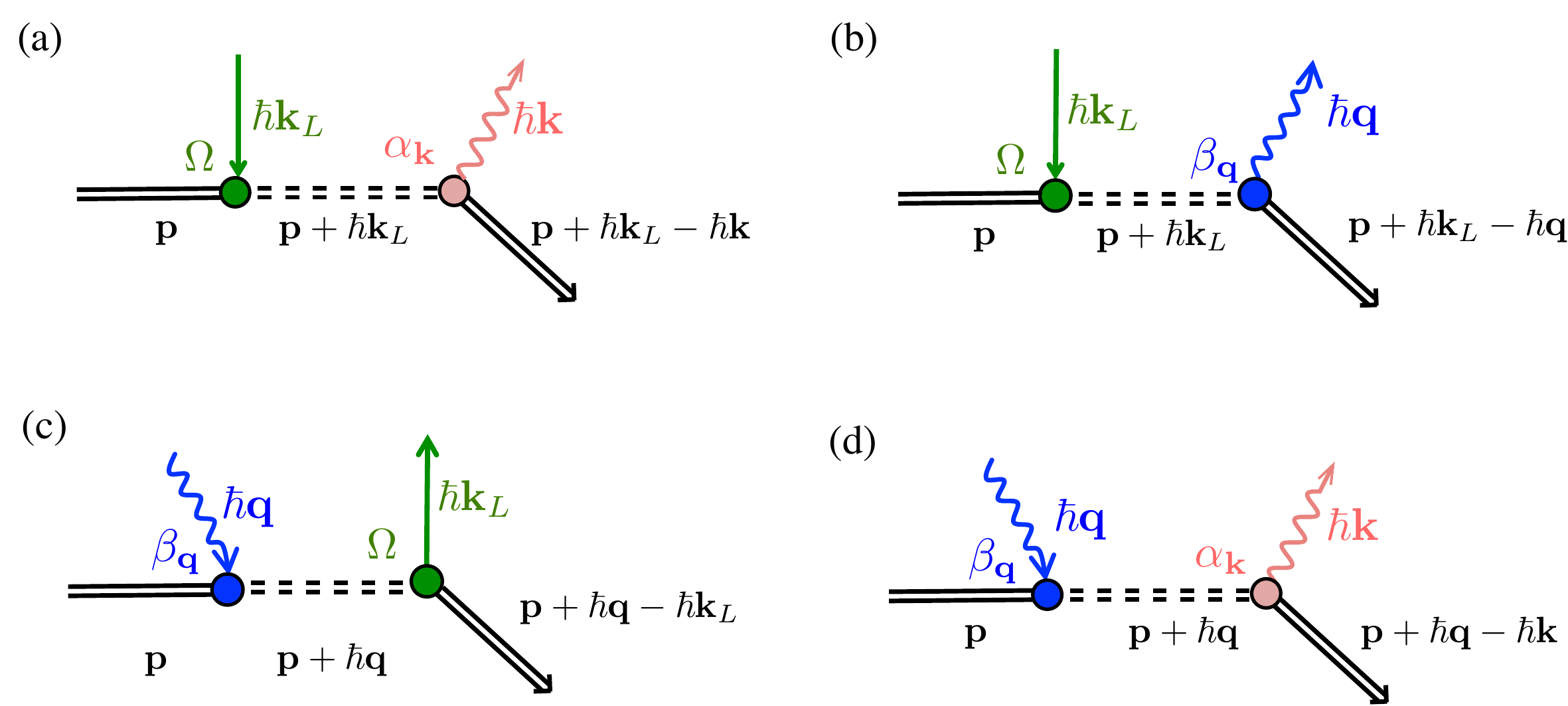}
\caption{The diagrammatic representation of the scattering amplitude for four possible processes associated with transitions out of the initial state $\ket{g,\bp}$ into final atomic states with a change in the ground-state momentum. (a) Scattering process that absorbs a laser photon and spontaneously decays into the bath modes. (b) Scattering process that absorbs  a laser photon and emits a system photon. (c) Scattering process that absorbs a system photon and scatters back into the laser mode. (d) Scattering process that absorbs a system photon and scatters into the bath modes.
 }
\label{fig:transitions}
\end{center}
\end{figure}

Our analysis makes use of a Born approximation, which assumes that the atomic momentum thermalizes (due to laser cooling) after each emission or absorption event of system photons as shown in Appendix A, leading to no correlations between the motional distribution and the system photons.  Thus we take the steady state motional distribution to be
\begin{equation}
\Pi(\bp) \mathrm{d}^3 \bp=\left(\frac{\beta}{2 \pi m}\right)^{\frac{3}{2}} e^{-\beta \abs{\bp}^2/2m} \mathrm{d}^3\bp
\label{Boltzmann}
\end{equation}
with a temperature $k_B T\approx \frac{\hbar}{2}\frac{\bar{\Delta}_L^2+\Gamma^2/4}{\abs{\bar{\Delta}_L}}$ set by laser cooling. 
In this approximation we can use Eq.~(\ref{Boltzmann}) to integrate over the atomic motion, and get an average rate for the thermalization of system photons. We first focus on the two processes involving system photon-laser photon scattering [see Fig.~\ref{fig:one}(b)-~\ref{fig:one}(c) and Fig.~\ref{fig:transitions}(b)-~\ref{fig:transitions}(c)]. 
 This steady-state distribution of the atoms effectively averages out the phase factor $ e^{i(\bk_L-\bq)\cdot\br} $ in the atom-light coupling $V_{ASL}(t)$ (see Appendixes A and B).  As a result, we can neglect coherent driving of the system photons, and the steady-state density matrix of the system photons is diagonal in the photon number basis.  The long-time dynamics is then governed by incoherent transitions between photon number sectors with rates computed below.

According to the SC-FGR and after integrating over the atomic momentum, the total rate to emit a system photon is given by $|\bra{n_\bq+1}a^{\dagger}_\bq\ket{n_\bq}|^2 \Lambda^+_{\bq}=(n_\bq+1)\Lambda^+_{\bq}$, with a laser-mediated single photon-emission rate given by the SC-FGR formula:
\begin{align}
\Lambda^+_{\bq,L}&=\int \mathrm{d}^3 \bp\Pi(\bp) |R^+_\bq(\bp)|^2 \delta_{\gamma(\bp) + \gamma(\bp+ \bk_L-\bq)}\left( \Delta E_{gg}(\bq,\bp) \right)
\label{lambda_L}
\end{align}
The decay rates of the initial and final momentum states are summed together in the broadened $\delta$ function because we evaluate the propagator in Eq.~(\ref{propagator}) at the on-shell energy of the intermediate state, which includes the decay rate. By analogy to Eq.~(\ref{gammap}), we refer to Eq.~(\ref{lambda_L}) as an example of self-consistent Fermi's golden rule because of the appearance of the decay-broadened $\delta$ function $\delta_{\gamma(\bp) + \gamma(\bp+ \bk_L-\bq)}$. 

To evaluate Eq.~(\ref{lambda_L}), we will use the high-temperature approximation discussed in Sec. II. The primary reason we introduced the SC-FGR is to quantitatively determine the regime of validity of this approximation. In particular, we find the condition
\begin{align}
{\gamma(\bp)} \ll \sqrt{\frac{k_B T}{m}}|\bk_L-\bq|,
\label{highT}
\end{align}
for which the decay-broadened $\delta$ function in Eq.~(\ref{lambda_L}) can be approximated by a true $\delta$ function since the integral over atomic momentum is much wider in energy than the decay broadening. More intuitively, this high-temperature limit can be interpreted as the condition that the momentum transfer to the atom is well defined, which requires that the Doppler broadening $v_{\rm th} \delta q$ associated with the thermal velocity $v_{\rm th}=\sqrt{k_B T/m}$ and atomic momentum transfer $\delta q=|\bk_L-\bq|$  is much greater than the motional decay rate of  the ground states $\gamma(\bp)$. With the additional approximation $\gamma(\bp) \ll \Gamma$ we find
\begin{align}
\Lambda^+_{\bq,L} \approx \frac{\sqrt{2\pi \beta m} \Omega^2|\alpha_{\bq}|^2  e^{-\beta p_{0}^2/2m}}{|\bk_L-\bq|\left[(\bar{\Delta}_L-\frac{p_{0} \hat{\bm{n}} \cdot \bk_L}{m})^2+\frac{\Gamma^2}{4}\right]},
\end{align}
where $p_{0}$ is the magnitude of atomic momentum, satisfying the energy conservation condition 
\begin{align}
\frac{p_{0} \abs{\bk_L- \bq}}{m}+\frac{\hbar (\bk_L-\bq)^2}{2m}+\omega_{\bq}-\omega_L=0
\end{align}
[the $\delta$-function argument in Eq.~(\ref{lambda_L})]
and $\hat{\bm{n}}=\frac{\bk_L-\bq}{\abs{\bk_L-\bq}}$ is the unit vector along the change between the initial and final momentum.

Similarly, one can find the total rate to absorb a system photon through the process that a system photon is first absorbed by the atom and then scattered back into the laser field [Fig.\ref{fig:transitions}(c)], $n_\bq \Lambda^-_{\bq,L}$, with a laser-mediated single-photon absorption rate in the high-temperature limit given by
\begin{align}
&\Lambda^-_{\bq,L} \approx 
 \frac{\sqrt{2\pi \beta m} \Omega^2|\alpha_{\bq}|^2  e^{-\beta p_0{'}^2/2m}}{|\bk_L-\bq|\left[(\bar{\Delta}_\bq - \frac{p_0{'} \hat{\bm{n}} \cdot \bq}{m})^2+\frac{\Gamma^2}{4}\right]}.
\label{lambda-L}
\end{align}
Here $p_0{'}$ is the magnitude of the atomic momentum satisfying the energy conservation condition 
\begin{align}
\frac{p_0{'} \abs{\bk_L- \bq}}{m}-\frac{\hbar (\bk_L-\bq)^2}{2m}+\omega_{\bq}-\omega_L=0.
\end{align}

If we consider the equilibration between these two processes only, the detailed balance condition reproduces the result of Eq.~(\ref{GrandL}), which we  motivated in Sec. II:
\begin{align}
\frac{\bar{n}_\bq+1}{\bar{n}_\bq}=\frac{\Lambda^-_{\bq,L}}{\Lambda^+_{\bq,L}}=\frac{e^{-\beta\frac{ p_0{'}^2}{2m}}}{e^{-\beta\frac{ p_{0}^2}{2m}}}=e^{-\beta \frac{p_0{'}^2-p^2_{0}}{2m}}=e^{\beta \hbar(\omega_{\bq} -\omega_L)}.
\label{Grand}
\end{align}
Here we have applied the equalities $ \frac{p_0{'}^2}{2m}-\frac{p_{0}^2}{2m}=\hbar \omega_L-\hbar \omega_{\bq}$, and $\bar{\Delta}_\bq - \frac{p_0{'} \hat{\bm{n}} \cdot \bq}{m}=\bar{\Delta}_L - \frac{p_{0} \hat{\bm{n}} \cdot \bk_L}{m}$. For $\omega_{\bq} > \omega_L$, we will have $\bar{n}_\bq=\frac{1}{e^{\beta \hbar(\omega_{\bq} -\omega_L)}-1}$, which corresponds to a Bose grand canonical distribution with temperature $\beta$ and an effective chemical potential $\hbar \omega_L$. For $\omega_{\bq} < \omega_L$, Eq.~(\ref{Grand}) suggests the onset of gain---higher photon numbers become ever more probable. A full treatment of that regime is beyond the present work. However, the use of a cavity can modify the system photon density-of-states to prevent gain from contributing to the dynamics (e.g., by setting the relevant cavity resonant frequencies higher than the laser frequency). 
As shown in Fig.~\ref{fig:one}(d) and below, once loss is properly taken into account, there is only a finite range of frequencies where the gain exceeds the loss.

\subsection{Accounting for additional photon loss mechanisms}
The detailed balance condition we found in Eq.~(\ref{Grand}) will be modified by system photon loss associated with scattering into the bath modes [Fig.~\ref{fig:transitions}(d)] or via the cavity mirrors. Considering first the loss into bath modes using SC-FGR, we find the overall scattering rate $n_\bq \Lambda^-_{\bq,B}$ in the high-temperature approximation, and neglecting the Doppler shift relative to the detuning $\bar{\Delta}_\bq$,
\begin{align}
\Lambda^-_{\bq,B} \equiv \int \mathrm{d}^3 \bp \frac{ \Pi(\bp)|\alpha_{\bq}|^2  \Gamma}{(\bar{\Delta}_\bq - \frac{\bp \cdot \bq}{m})^2+\frac{\Gamma^2}{4}} \approx \frac{|\alpha_{\bq}|^2  \Gamma}{\bar{\Delta}_\bq^2+\frac{\Gamma^2}{4}}.
\label{lambda-B}
\end{align}
Neglecting the Doppler shift in $\Lambda_{\bq,L}^-$, we find the modified detailed balance condition
\begin{align}
&\frac{\bar{n}_\bq+1}{\bar{n}_\bq}=\frac{\Lambda^-_{\bq,L}+\Lambda^-_{\bq,B}}{\Lambda^+_{\bq,L}} \notag\\
&\approx e^{\beta\hbar(\omega_{\bq}-\omega_L)}+\frac{\Gamma|\bk_L-\bq|}{\Omega^2\sqrt{2 \pi \beta m}}e^{\frac{\beta}{2m}\left(-\frac{m (\omega_{\bq}-\omega_L)}{|\bk_L-\bq|}-\frac{\hbar|\bk_L-\bq|}{2}\right)^2}.
\label{DetailedBalance}
\end{align}
If the scattering loss rate is small, $\Lambda^-_{\bq,B} \ll \Lambda^-_{\bq,L}$, one can treat the effect of loss as a small correction to Eq.~(\ref{GrandL}) and identify an effective temperature for the cavity mode $k_B T_{\rm eff}=\beta_{\rm eff}^{-1}$ and an observed shift to the chemical potential $\delta \mu$ according to the modified condition Eq.~(\ref{DetailedBalance}):
\begin{align}
\beta_{\rm eff}(\hbar \omega_{\bq}- \hbar\omega_L+\delta \mu) = \ln\Big(\frac{\bar{n}_\bq+1}{\bar{n}_\bq}\Big).
\label{Teff}
\end{align}
Here the observed shift in the chemical potential $\delta \mu$, typically much smaller than the atomic temperature, is formally defined such that $\hbar \omega_{\bq}=\hbar \omega_L-\delta \mu$ is the transition frequency from equilibrium to gain,
\begin{align}
\frac{\bar{n}_\bq+1}{\bar{n}_\bq} = 1=  e^{-\beta\delta \mu}+\frac{\Gamma|\bk_L-\bq|}{\Omega^2\sqrt{2 \pi \beta m}}e^{\frac{\beta}{2m}\left(\frac{m \delta \mu}{\hbar|\bk_L-\bq|}-\frac{\hbar|\bk_L-\bq|}{2}\right)^2},
\label{omegao}
\end{align}
where $\delta \mu$ and $\beta_{\rm eff}$ can potentially depend on $\bq$.  In Sec.~\ref{sec:GCE}, we determine the conditions under which this $\bq$-dependence can be neglected, in which case a single temperature and chemical potential [yellow region in Fig.~\ref{fig:one}(d)] describe the relevant system modes over a wide range of frequencies.  

To aid concreteness in the remaining discussion,  we  focus on a Fabry-Perot cavity design as illustrated  in Fig.~\ref{fig:one}(a). In the regime of interest, the dependence of $\bq$ on $\omega_{\bq}$ is weak enough to neglect, and the only $\bq$ dependence that remains is in its angle relative to $\bk_L$. For a Fabry-Perot cavity, the system modes are nearly colinear and the angle dependence of $\delta \mu$ and $\beta_{\rm eff}$ can also be neglected. Equation~(\ref{omegao}) does not always have solutions; for a given $\bq$ direction, there is a critical laser Rabi frequency $\Omega_c$ below which there are no solutions. This value determines the minimum power required to observe grand canonical ensemble behavior with a well-defined chemical potential, setting the bottom of the gain region in Fig.~\ref{fig:one}(d). Above this critical power, there are always two solutions, which determine the left and right boundaries of the gain region in Fig.~\ref{fig:one}(d).  Finally, we use the left boundary as the definition of $\delta \mu$.  

To thermalize close to a GCE with $\beta_{\rm eff} \approx \beta$, we require the coefficient of the second term in Eq.~(\ref{omegao}) to be much less than $1$.  This is equivalent to the condition $\Lambda^-_{\bq,B} \ll \Lambda^-_{\bq,L}$, which requires 
\begin{align}
 \sqrt{\frac{k_B T}{m}} {|\bk_L-\bq|} \ll \frac{\Omega^2}{\Gamma}.
\label{Smallloss}
\end{align}
The high-temperature limit has already set a constraint on the left-hand side of this equation. Combining the inequality Eq.~(\ref{Smallloss}) with the high-temperature limit Eq.~(\ref{highT}), using the explicit form of $\gamma(\bq)$ Eq.~(\ref{gammap}) and the approximation $\abs{\bk_L-\bq} \approx \abs{\bk_L}\sqrt{2(1- \cos \theta)}$, the condition for the system photon to thermalize close to a grand canonical ensemble is
\begin{align}
\frac{\Gamma^4}{\left(\bar{\Delta}_L^2+\frac{\Gamma^2}{4} \right)^2} \ll \frac{\Gamma^2 \left(\bar{\Delta}_L^2+\frac{\Gamma^2}{4}\right)E_r(\bk_L)}{\Omega^4  \abs{\bar{\Delta}_L}\hbar} (1- \cos \theta) \ll 1.
\label{Requirement}
\end{align}

The above inequalities can be satisfied in the low-excitation limit with 
\begin{align}
\Gamma \ll \Omega \ll |\bar{\Delta}_L|, 
\label{thermalregime}
\end{align}
with a finite angle $\theta$, and assuming $E_r(\bk_L)/\hbar \lesssim \Gamma$ as one typically finds for laser cooling transitions. This condition can be understood intuitively: 
In order for the laser photon-system photon scattering rate to dominate over the system photon scattering loss, one needs to  increase the pump intensity until $\Gamma \ll \Omega$; a large detuning $\Omega \ll |\bar{\Delta}_L|$ is then required to stay in the low-excitation limit, leading to a higher atomic temperature than the standard detuning case with $\bar{\Delta}_L\approx -\Gamma/2$. For strong pump intensity, $\Omega > |\bar{\Delta}_L|,\Gamma$, we would need to revisit the problem nonperturbatively in $\Omega/\abs{\bar{\Delta}_L+i \Gamma}$, which may remove this high-temperature tradeoff.

To account for a finite cavity loss $\kappa_\bq$, the detailed balance condition is furthered modified as
\begin{align}
\frac{\bar{n}_\bq+1}{\bar{n}_\bq}=\frac{\Lambda^-_{\bq,L}+\Lambda^-_{\bq,B}+\kappa_\bq}{\Lambda^+_{\bq,L}}.
\label{DetailedBalanceKappa}
\end{align}
The cavity loss will increase the critical laser Rabi frequency $\Omega_c$, and further modify the observed shift in chemical potential  $\delta \mu$ and the effective temperature $T_{\rm eff}$. 
A cavity with small enough linewidth, i.e., with cavity loss rate much slower than the system photon emission and absorption rates, is thus also required to achieve a grand canonical distribution of system photons. In practice, one can increase the optical depth by adding more atoms into the ensemble to overcome the contribution from cavity loss since it is independent of the number of atoms. We neglect the cavity loss in our calculations for Fig.~\ref{fig:one}(d) and the next section.

\section{Realizing the Grand Canonical Ensemble limit}\label{sec:GCE}

\begin{figure}[htbp!]
\begin{center}
\includegraphics[width=0.45 \textwidth]{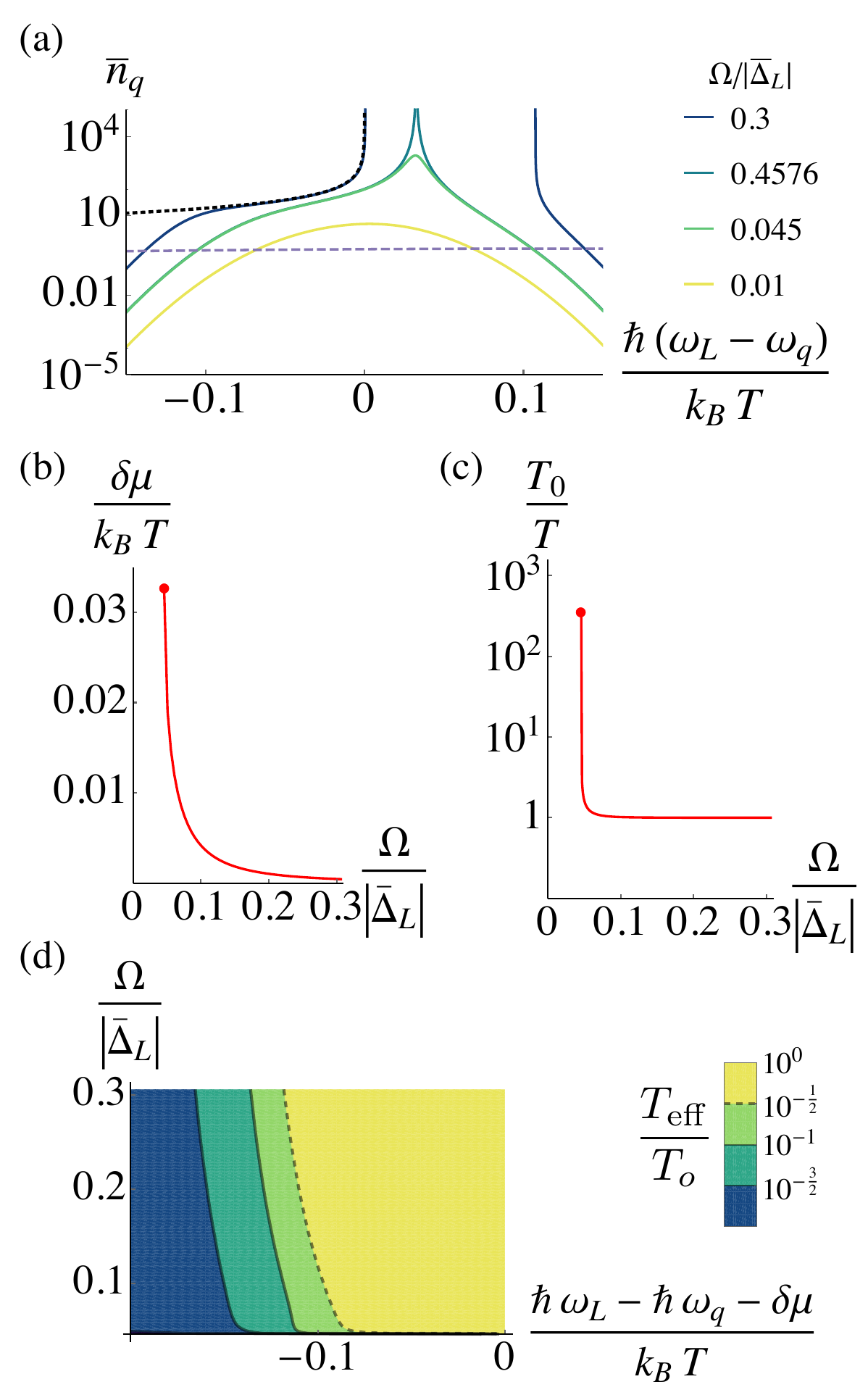}
\caption{Grand canonical ensemble realization for a tunable-frequency single-mode cavity with a Yb gas. (a) Equilibrium system photon occupation number ($\bar{n}_{\bq}$) as a function of the laser detuning from the system photon frequency ($\omega_L-\omega_{\bq}$). The ideal grand canonical ensemble result is plotted as the dotted black line, the standard detuning with $\bar{\Delta}_L\approx-\frac{\Gamma}{2}$ and $\Omega=0.15 \Gamma$ is shown as the purple dashed line, and for large detuning with $\bar{\Delta}_L \approx - 157 \Gamma$ and varying $\Omega$ are shown with colored solid lines. (b) The observed shift in the equilibrium-to-gain chemical potential ($\delta \mu$) as a function of the laser intensity ($\Omega$) with $\bar{\Delta}_L \approx - 157 \Gamma$. (c) The effective temperature at the equilibrium-to-gain transition ($T_o$) as a function of the laser intensity ($\Omega$) with $\bar{\Delta}_L \approx - 157 \Gamma$. (d) The ratio between the effective temperature $T_{\rm eff}$ and $T_o$ as a function of the shifted laser detuning from the system photon frequency ($\omega_L-\omega_{\bq}-\delta \mu/\hbar$) and the laser intensity ($\Omega$) with $\bar{\Delta}_L \approx - 157 \Gamma$. The $y$-axis has a lower cutoff $\Omega_c/\abs{\bar{\Delta}_L}=0.04576$ which also corresponds to the end point (dot) in Figs.~\ref{fig:Teff}(b)-\ref{fig:Teff}(c). In these plots we assume $\abs{\bk_L-\bq} \approx \sqrt{2}\abs{\bk_L}$ and take the parameters for the Yb intercombination transition to ${}^{3}P_1$, $\omega_A/2\pi =$ 539 THz, $\Gamma/2\pi=$ 180 kHz, $E_{r,\bk_L}/h=$ 3.74 kHz.
}
\label{fig:Teff}
\end{center}
\end{figure}

We now numerically study the results of the modified detailed balance equation (Eq.~(\ref{DetailedBalance})) to verify our previous analysis and characterize in which regimes the photon steady state is described by a single temperature and chemical potential. The equilibrium system photon occupation number for several conditions is shown in Fig.~\ref{fig:Teff}(a). In the standard Doppler cooling case, $\bar{\Delta}_L \simeq -\Gamma/2$, the mean system photon number is always small due to the rapid scattering of system photons into bath modes, and the grand-canonical-like distribution cannot be achieved, as suggested in Sec.~\ref{sec:SCFGR} B. [See the dashed line in Fig.~\ref{fig:Teff}(a).]

On the other hand, in the large detuning regime, the photon occupation number at negative cavity detuning may approach the distribution for an ideal grand canonical ensemble, as described in Eq.~(\ref{GrandL}). This occurs for a laser Rabi frequency larger than the critical value $\Omega_c$ [blue and orange curves in Fig.~\ref{fig:Teff}(a)] in the negative detuning regime. We remark that the chemical potential is shifted from the laser frequency slightly by an amount $\delta \mu$, as discussed in Sec.~\ref{sec:SCFGR} B.  An example plot  $\delta \mu$ as a function of $\Omega$ is shown in Fig.~\ref{fig:Teff}(b), where the critical end point at $\Omega_c$ is indicated as a dot. When the chemical potential exceeds the single-photon energy, gain is expected, leading to diverging photon numbers as seen in Fig.~\ref{fig:Teff}(a), which defines the green region in Fig.~\ref{fig:one}(d). By further increasing the laser frequency beyond the gain region to positive cavity detuning, the photon occupation number becomes finite again, indicating quasithermal behavior distinct from the grand canonical description. 
For laser intensity less than the critical value [green and red curves in Fig.~\ref{fig:Teff}(a)], the photon occupation number never diverges and there is gain-free region. 

In the large detuning regime, we characterize the steady-state behavior of system photons as a function of laser frequency and intensity by quantifying the degree to which $T_{\rm eff}$ is independent of $\bq$. For a reference temperature we use the value of the effective temperature at the equilibrium-to-gain transition ($T_o$), defined as $T_{\rm eff}$ when $\omega_\bq =\omega_L-\delta \mu/\hbar$, which is shown in Fig.~\ref{fig:Teff}(c) as a function of the laser intensity. Above $\Omega_c$, $T_o$ quickly approaches the atomic temperature. The calculated ratio $T_{\rm eff}/ T_o$ as a function of mode frequency and laser intensity is shown in Fig.~\ref{fig:Teff}(d). This ratio quantifies the degree to which the system photons can be well characterized by a single chemical potential and temperature, with a ratio of 1  over a large range of $\bq$ indicating perfect thermalization. We choose the yellow region in Fig.~\ref{fig:Teff}(d)---identified as the grand canonical ensemble (GCE) region in Fig.~\ref{fig:one}(d) as well---by defining the condition that $-1/2 \leq \log_{10}\left(\frac{T_{\rm eff}}{T_o}\right) \leq 0$. Outside the gain and GCE region, i.e., at low laser powers or large detunings, photon loss prevents detailed balance with the atomic motion, and only quasithermal light is expected [blue region in Fig.~\ref{fig:one}(d)].

\section{Outlook}
\label{sec:outlook}
We have identified an application of Doppler cooling of atoms by considering the steady state of the re-emitted light and showed this light can be described as a grand canonical ensemble with a laser-controlled chemical potential and a temperature set by the atomic motional temperature. Our analysis offers a framework to study the behavior of optically thick ensembles. Looking forward, the simplicity of our approach---using an ensemble of two-level atoms contained within an optical cavity, and maintaining a balance between optical depth and transparency---will admit a variety of extensions and expansions. For example, we can examine sub-Doppler regimes, cavity-assisted cooling, and related phenomena. An immediate consequence of this paper is that 
Bose condensation of noninteracting photons via laser cooling of atoms inside a multimode cavity should be possible; we defer the details of this for a later work.  With a fully microscopic treatment and thermodynamic detailed balance arguments, our approach can be directly applied to more exotic interacting photonic systems \cite{Hafezi2015}. For example, adding synthetic gauge fields to the problem would map the cavity system to an interacting quantum Hall system. Another promising future direction will be studying Rydberg-polariton thermalization with laser-cooled Rydberg atoms working in the electromagnetically induced transparency (EIT) regime. This may provide a cavity-free setting for observing equilibrium behavior of interacting photons, where intriguing many-body phenomena can arise.

\begin{acknowledgments}
We thank A. V. Gorshkov, E. A. Goldschmidt, V. Vuleti\'{c}, M. Weitz, H. Carmichael, and S. Ragole for helpful discussions. Funding is provided by NSF Physics Frontier Center, PFC @ JQI.
\end{acknowledgments}

\appendix

\section{The Master Equation}
\label{sec:mastereq}

The rapid decay of the lossy bath modes $b_\bk$ leads to dissipative effects for the atoms (e.g., excited-state decay) and the system photons $a_{\bq}$ (via scattering into bath modes). We describe the resulting dissipative dynamics with a Lindblad-form master equation.
Each Lindblad-form damping term has a jump operator $\hat{c}_j$ and an associated rate $r_j$. Formally, 
the master equation for a density matrix $\rho_{\rm T}$ describing the atoms and the system photons is~\cite{Carmichael1993}
\begin{align}
&\frac{\mathrm{d} \rho_{\rm T}}{\mathrm{d}t}=\frac{1}{i\hbar}(H_{\rm eff}  \rho_{\rm T} -  \rho_{\rm T} H_{\rm eff}^{\dagger}) + \sum_j r_j \hat{c}_j \rho_{\rm T} \hat{c}_j^{\dagger}, \notag\\
&H_{\rm eff}=H_{\rm T} - i \hbar \sum_j \frac{r_j}{2}\hat{c}^{\dagger}_j \hat{c}_j,
\label{MasterJ}
\end{align}
where $H_{\rm T}$ is the combined system and atom Hamiltonian.

We now identify the jump operators and corresponding rates for atomic states due to bath modes based on the above analysis. Due to spontaneous decay into bath modes, for each bath mode $b_\bk$, we have an excited-state jump operator $\hat{c}^e_\bk$ with a jump rate $r^e_\bk$:
\begin{align}
\hat{c}^e_\bk&=e^{-i\bk\cdot \br}\ket{g}\bra{e}, \notag\\ r^e_\bk(\bp)&= \frac{2 \pi}{\hbar} \abs{\hbar \beta_\bk}^2  \delta(\Delta E_{eg}(\bk,\bp)).
\label{cere}
\end{align}
The sum of all jump rates leads to an overall decay rate of the excited state, $\sum_\bk r^e_\bk=\Gamma$, as expected. The second-order transitions due to laser-bath scattering with modes $b_\bk$ will also lead to ground state--ground state jump operators $\hat{c}^g_\bk$ and jump rates $r^g_\bk$,
\begin{align}
\hat{c}^g_\bk&=e^{i(\bk_L-\bk)\cdot \br}\ket{g}\bra{g}, \notag\\ r^g_{\bk}(\bp)&=\frac{2 \pi}{\hbar} \abs{\hbar R_\bk(\bp)}^2 \delta(\Delta E_{gg}(\bk,\bp)),
\label{cgrg}
\end{align}
where $R_\bk(\bp)$ is given by Eq.~(\ref{Rbk}).
The sum of all jump rates leads to a laser-induced decay of the atomic ground state, $\sum_\bk r^g_{\bk}(\bp) = \gamma (\bp) $. Note that the groun-state decay rate is momentum dependent. The overall effect of the cavity decay of the system photons can be described as jump operators $\hat{c}_\bq=a_{\bq}$, and jump rates $\kappa_\bq$.

When applying the ground-state jump operators and rates, the lowest order effect in $\Omega$ of the bath modes $b_{\bk}$ has been included. To treat the overall problem consistently, we have to express the Hamiltonian to the lowest order in $\Omega$ for the system modes $a_{\bq}$ as well. Specifically, we have
\begin{align}
V_{ASL}(t)=& -\sum_\bq  a^{\dagger}_{\bq} e^{-i \omega_L t} e^{i (\bk_L-\bq) \cdot \br} \hbar R^+_\bq(\bp)\ket{g} \bra{g} \notag\\
& -\sum_\bq  a_{\bq} e^{i \omega_L t} e^{-i (\bk_L-\bq) \cdot \br} \hbar R^-_\bq(\bp) \ket{g} \bra{g},
\label{VQL(t)}
\end{align}
where $R^+_\bq(\bp)$ and $R^-_\bq(\bp)$ are given by Eqs.~(\ref{R+bq}) and ~(\ref{R-bq}).

We can then write down the effective Hamiltonian after integrating out the lossy bath modes $b_\bk$ as
\begin{align}
H_{\rm eff}=&\hbar\left(\omega_A - i \frac{\Gamma}{2}\right) \ket{e}\bra{e} +\frac{\bp^2}{2m}-\frac{i \hbar  \gamma(\bp)}{2} \ket{g}\bra{g}\notag\\
&+\sum_\bq \hbar \left(\omega_{\bq}-i\frac{\kappa_\bq}{2}\right) a_{\bq}^\dagger a_{\bq} \notag\\
& -\sum_\bq \hbar \alpha_{\bq}  \left( a_{\bq} e^{i \bq \cdot \br} \ket{e} \bra{g}+ \ket{g} \bra{e} a_{\bq}^{\dagger}e^{-i \bq \cdot \br}\right) \notag\\
& -\sum_\bq  a^{\dagger}_{\bq} e^{-i \omega_L t} e^{i (\bk_L-\bq) \cdot \br} \hbar R^+_\bq(\bp)\ket{g} \bra{g} \notag\\
& -\sum_\bq  a_{\bq} e^{i \omega_L t} e^{-i (\bk_L-\bq) \cdot \br} \hbar R^-_\bq(\bp) \ket{g} \bra{g}.
\label{Heff}
\end{align}
From the $V_{ASL}(t)$ term in the effective Hamiltonian, we see explicitly that the laser mediates time-dependent (parametric) coupling between the system photons and the atomic ground state. 
We will study how the atomic motional state evolves later in this section, which determines the dynamics of the bath. In Appendix B we will find the emission and absorption rates of the system photons on top of the rapid thermalization of atoms due to laser and bath modes.

Working from Eq.~(\ref{MasterJ}), we write the full  master equation for the atoms and system photons:
\begin{align}
\frac{\mathrm{d} \rho_{\rm T}}{\mathrm{d}t}&=\frac{1}{i\hbar}(H_{\rm eff} \rho_{\rm T} - \rho_{\rm T} H_{\rm eff}^{\dagger}) + \sum_{\bq} \kappa_{\bq} a_{\bq} \rho_{\rm T} a^{\dagger}_{\bq}.\notag\\ 
&+ \sum_\bk \left\{\frac{r^e_\bk}{2}, e^{-i\bk\cdot \br} \ket{g} \bra{e} \rho_{\rm T} \ket{e}\bra{g}  e^{i\bk\cdot \br} \right\} \notag\\ 
&+ \sum_\bk \left\{\frac{r^g_\bk}{2}, e^{i(\bk_L-\bk)\cdot \br} \ket{g} \bra{g} \rho_{\rm T} \ket{g}\bra{g}  e^{-i(\bk_L-\bk)\cdot \br} \right\}
\label{Master}
\end{align}
Since the jump rates depend on the atomic momentum, we treat them as operators and symmetrize them around $\hat{c}_j \rho_{\rm T} \hat{c}^{\dagger}_j$ terms using the anticommutator $\{,\}$.

The system density matrix now describes the atoms and the system photons $a_{\bq}$ only, and can be further separated as $\rho_{\rm T}=\rho_{\rm ee}\ket{e} \bra{e}+\rho_{\rm gg}\ket{g} \bra{g}+\rho_{\rm eg}\ket{e} \bra{g}+\rho_{\rm ge}\ket{g} \bra{e}$, where $\rho_{ij}$ is an operator acting in the atomic momentum and photon number Hilbert space. According to the master equation, the excited component $\rho_{\rm ee}$ is rapidly decaying with a rate $\Gamma$ while the off-diagonal terms $\rho_{\rm eg}$, $\rho_{\rm ge}$ dephase with a rate $\Gamma/2+\gamma(\bp)/2 \approx \Gamma/2$.

We now focus on the ground state component $\rho_{\rm gg}$ only. We start with the simplest case that the overall space is one dimensional (1D) along the $\hat{x}$ axis. In the limit $(\bk_L-\bk) \cdot \br \ll 1$, we can make a Lamb-Dicke approximation because the distance the atom moves during scattering events is much shorter than the wavelength of the photons. This leads to the approximate expression
\begin{align}
    &e^{i  (\bk_L-\bk) \cdot \br} \rho_{gg} e^{-i (\bk_L-\bk) \cdot \br} \approx \rho_{gg}+i\comm{(\bk_L-\bk) \cdot \br}{{\rho_{gg}}}\notag\\ 
&-\frac{1}{2}\comm{(\bk_L-\bk) \cdot \br}{\comm{(\bk_L-\bk) \cdot \br}{\rho_{gg}}}.
    \label{LambDicke}
\end{align}
Applying the first term in Eq.~(\ref{LambDicke}) to Eq.~(\ref{Master}) produces $\pbrac{\frac{\gamma(\bp)}{2}}{\rho_{\rm gg}}$, which cancels with $-\pbrac{\frac{\gamma(\bp)}{2}}{\rho_{\rm gg}}$ from the imaginary part of the effective Hamiltonian in ground state, $-\frac{i \hbar \gamma(\bp)}{2} \ket{g}\bra{g}$. 

The second term in Eq.~(\ref{LambDicke}) is imaginary and will lead to an effective force in Eq.~(\ref{Master}). Assuming the recoil effect is small as in the usual Doppler cooling scheme, the relevant bath modes are photons with momentum about the same magnitude as $\hbar k_L$ but nearly isotropic. Therefore, we have $\sum_\bk \bk \abs{\beta_\bk^2} \delta (\Delta E_{gg}(\bk,\bp)) \approx 0$. For $\bk_L=k_L \hat{x}$, the second term leads to $\sum_\bk i k_L \comm{x}{\left\{ r^g_\bk, \rho_{\rm gg} \right\}} \approx \frac{i k_L \Omega^2 \Gamma}{\bar{\Delta}_L^2+\Gamma^2/4} \left( \comm{x}{\rho_{\rm gg}}+\frac{\bar{\Delta}_L k_L}{(\bar{\Delta}_L^2+\Gamma^2/4)m}\comm{x}{\pbrac{p_x}{\rho_{\rm gg}}}\right)$. The approximation follows by assuming the Doppler shift and the recoil shift are much smaller than $\bar{\Delta}_L$ or $\Gamma$. The part proportional to $\comm{x}{\rho_{\rm gg}}$ corresponds to a dc force $\frac{\Omega^2 \Gamma}{\bar{\Delta}_L^2+\Gamma^2/4} \hbar \bk_L$, and the $\comm{x}{\pbrac{p_x}{\rho_{\rm gg}}}$ part will create a velocity-dependent damping term for $p_x$ with a rate $\frac{-2 \hbar \Omega^2 \Gamma \bar{\Delta}_L k^2_L}{(\bar{\Delta}_L^2+\Gamma^2/4)^2 m}$. Note that the rate is positive with a negative detuning $\bar{\Delta}_L$ appropriate for laser cooling.

We then examine the third term in Eq.~(\ref{LambDicke}), $-\frac{1}{2}\comm{(\bk_L-\bk) \cdot \br}{\comm{(\bk_L-\bk) \cdot \br}{\rho_{\rm gg}}}$. Recall that we are considering the 1D case, we have $\sum_\bk (\bk_L-\bk)^2 \frac{2 \pi}{\hbar}  \abs{\hbar\beta_\bk^2}\delta( \Delta E_{gg}(\bk,\bp)) \approx 2 k_L^2 \Gamma$. We have again neglected the recoil effect for the relevant bath modes. The leading order contribution from this term to Eq.~(\ref{Master}) is $-\frac{\Omega^2 \Gamma k_L^2}{\bar{\Delta}_L^2+\Gamma^2/4} \comm{x}{\comm{x}{\rho_{\rm gg}}}$, which corresponds to a diffusion term for the atomic momentum.

The master equation in 1D for the atoms after elimination of the excited state now reads
\begin{align}
\frac{\mathrm{d} \rho_{\rm gg}}{\mathrm{d}t}&=
\frac{1}{i \hbar}\comm{H_{\rm  T}-i\hbar\sum_{\bq}\frac{\kappa_{\bq}}{2}a^{\dagger}_{\bq}a_{\bq}}{\rho_{\rm gg}}\notag\\&
+\frac{1}{i \hbar}\comm{- F_0(\bk_L) x}{\rho_{\rm gg}}-\frac{i \zeta}{\hbar}\comm{x}{\pbrac{p_x}{\rho_{\rm gg}}_+}\notag\\&
-\frac{2 m \zeta k_B T}{\hbar^2}\comm{x}{\comm{x}{\rho_{\rm gg}}}+ \sum_{\bq} \kappa_{\bq} a_{\bq} \rho_{\rm gg} a^{\dagger}_{\bq}.
\label{QBM}
\end{align}
Here $H_{\rm  T}=\frac{\bp^2}{2m}+\hbar \omega_A \ket{e}\bra{e}+ H_S+H_{AS}+V_{ASL}(t).$ The above equation is in fact the master equation for quantum Brownian motion theory with a damping constant $\zeta=\frac{ \hbar \Omega^2 \Gamma \abs{\bar{\Delta}_L} k^2_L}{(\bar{\Delta}_L^2+\Gamma^2/4)^2 m}$ and temperature $k_B T=\frac{\hbar}{2}\frac{\bar{\Delta}_L^2+\Gamma^2/4}{\abs{\bar{\Delta}_L}}$ given by the momentum diffusion up to a DC force term $F_0(\bk_L)=\frac{\hbar k_L \Omega^2 \Gamma}{\bar{\Delta}_L^2+\Gamma^2/4}$. The dc force term (a drift) can be compensated for by another dc force term or by including a counter-propagating laser. This master equation can be easily generalized to three dimensions. In practice, the use of multiple laser beams will remove the drift term and recover the standard Doppler cooling theory of two-level atoms in the low-excitation limit.

\section{Self-Consistent Fermi's Golden Rule using Time-Dependent Perturbation Theory}

To treat the dynamics of the atoms and the system photons self-consistently, we here calculate the transition rates associated with system photon emission and absorption using the quantum jump master equation and time-dependent perturbation theory. This complements---and indeed is equivalent to---the diagrammatic approach followed in the main text.
For simplicity, we consider the case of a single laser mode $\bk_L$, a single system photon mode $a_{\bq}$, and an initial state with a definite atomic momentum $\ket{\psi(t=0)}=\ket{n_{\bq}}\ket{g,\bp}$. Here $\{n_{\bq}\}$ denotes the Fock state of system photons. We also neglect cavity loss for the moment. Under the effective Hamiltonian, the time evolution of the unnormalized state is
\begin{align}
&\ket{\psi(t)} \approx  e^{- i z_{g_0} t} \ket{\psi(0)} \notag\\&-\frac{i}{\hbar} \sum_n \ket{n} \int_0^t \mathrm{d}t_a e^{-i z_n (t-t_a) - i z_{g_0} t_a} \bra{n} V(t_a) \ket{\psi(0)} \notag\\
&= e^{-i z_{g_0} t}\ket{g_0}+i \alpha_{\bq}   \sqrt{n_\bq} f_{e^-_\bq}(t)\ket{e^-_\bq} \notag\\ &+i R^+_\bq(\bp) \sqrt{n_\bq +1} f_{{g^+_{\bq,L}}}(t)\ket{g^+_{\bq,L}} \notag\\
&+i R^-_\bq(\bp) \sqrt{n_\bq} f_{g^-_{\bq,L}}(t) \ket{g^-_{\bq,L}}.
\label{psit}
\end{align}

According to the effective Hamiltonian, the evolution of each state has a complex frequency $z_i=\omega_i-i \gamma_i/2$; the real part corresponds to the state energy and the imaginary part denotes the damping. The short hand notations of the possible states are
\begin{align}
&\ket{g_0}=\ket{n_{\bq}}\ket{g,\bp},\notag\\
&\ket{g^+_{\bq,L}}=\ket{n_{\bq}+1}\ket{g,\bp+\hbar \bk_L-\hbar \bq},\notag\\
&\ket{g^-_{\bq,L}}=\ket{n_{\bq}-1}\ket{g,\bp+\hbar \bq-\hbar \bk_L},\notag\\
&\ket{e^-_\bq}=\ket{n_{\bq}-1}\ket{e,\bp+\hbar \bq},
\label{kets}
\end{align}
and their corresponding complex frequencies are defined as
\begin{align}
&z_{g_0}=\frac{\bp^2}{2m\hbar}+n_\bq \omega_{\bq}-\frac{i\gamma(\bp)}{2},\notag\\
&z_{g^+_{\bq,L}}=\frac{\abs{\bp+\hbar \bk_L-\hbar \bq}^2}{2m\hbar}+(n_\bq+1) \omega_{\bq}-\frac{i\gamma(\bp+\hbar \bk_L-\hbar \bq)}{2},\notag\\
&z_{g^-_{\bq,L}}=\frac{\abs{\bp+\hbar \bq-\hbar \bk_L}^2}{2m\hbar}+(n_\bq-1) \omega_{\bq}- \frac{i\gamma(\bp+\hbar \bq-\hbar \bk_L)}{2}, \notag\\
&z_{e^-_\bq}=\frac{\abs{\bp+\hbar \bq}^2}{2m\hbar}+(n_\bq-1) \omega_{\bq}+\omega_A- \frac{i\Gamma}{2}.
\label{zs}
\end{align}
Here the superscript $``+"$ again denotes emission of a system photon ($\ket{n_{\bq}} \rightarrow \ket{n_{\bq}+1}$) and $``-"$ for absorption ($\ket{n_{\bq}} \rightarrow \ket{n_{\bq}-1}$).

The time-dependent functions $f_i(t)$ are
\begin{align}
&f_{g^+_{\bq,L}}(t)=\int_0^{t} \mathrm{d}t_a e^{-i z_{g^+_{\bq,L}} (t-t_a)}  e^{-i\omega_L t_a-iz_{g_0} t_a},\\
&f_{g^-_{\bq,L}}(t)=\int_0^{t} \mathrm{d}t_a e^{-i z_{g^-_{\bq,L}} (t-t_a)}  e^{i\omega_L t_a-iz_{g_0} t_a},\\
&f_{e^-_\bq}(t)= \int_0^{t} \mathrm{d}t_a e^{-i z_{e^-_\bq} (t-t_a)}  e^{-iz_{g_0} t_a}.
\label{fs}
\end{align}
Compared to the usual time-dependent perturbation theory, the above time-dependent functions include the dissipative part of the Hamiltonian, and shall lead to modifications from the usual Fermi's golden rule.

First, since $\innerp{\psi(t)}{\psi(t)} \approx e^{-\gamma_\bp t}$, we find the jump time $t_J$ by solving $r = e^{-\gamma_\bp t_J}$, $0 \leq r \leq 1$, where $r$ is randomly distributed in $(0,1)$. The average total jump rate from the initial state is $\approx \gamma(\bp)$ up to a correction at order $V^2$. At the time of the jump, we need to evaluate the different possible jump outcomes, according to the un-normalized probability distributions $P_{j} \propto \mathcal{P}_{j} = \gamma_{j} \bra{\psi(t_J)} \hat{c}_{j}^\dag \hat{c}_{j} \ket{\psi(t_J)}/r$, and here we have included the $1/r$ factor to normalize the wave vector. We see
\begin{align}
\mathcal{P}_0(\bp)  &= \sum_\bk r^g_\bk(\bp) \bra{g} \hat{c}^{g \dag}_{\bk}\hat{c}^g_{\bk}\ket{g} \frac{ |e^{-i z_{g_0} t_J}|^2}{r}=\gamma(\bp) ,
\label{p0p}
\end{align}
\begin{align}
\mathcal{P}^+_{\bq ,L}(\bp)
=\gamma(\bp+\hbar \bk_L-\hbar \bq)|R^+_\bq(\bp)|^2(n_\bq+1)\frac{|f_{g^+_{\bq,L}}(t_J)|^2}{r},
\label{p+Lp}
\end{align}
\begin{align}
\mathcal{P}^-_{\bq ,L}(\bp)
=\gamma(\bp-\hbar \bk_L+\hbar \bq)|R^-_\bq(\bp)|^2n_\bq\frac{|f_{g^-_{\bq,L}}(t_J)|^2 }{r},
\label{p-Lp}
\end{align}
\begin{align}
\mathcal{P}^-_{\bq ,B}(\bp) 
=\Gamma|\alpha_{\bq}|^2n_\bq\frac{  |f_{e^-_\bq}(t_J)|^2 }{r}.
\label{p-Bp}
\end{align}
We interpret the possible jump outcomes as the following processes (see Fig.\ref{fig:transitions}): (a) $\mathcal{P}_0(\bp)$: scattering process that absorbs a laser photon and spontaneously decays into the bath modes;  (b) $\mathcal{P}^+_{\bq ,L}(\bp)$: scattering process that absorbs  a laser photon and emits a system photon; (c) $\mathcal{P}^-_{\bq ,L}(\bp)$: scattering process that absorbs a system photon and scatters back into the laser mode; (d) $\mathcal{P}^-_{\bq ,B}(\bp)$: scattering process that absorbs a system photon and scatters into the bath modes. The average rates for those processes will be evaluated in details.

Since $\ket{\psi(t_J)} \approx e^{- i z_{g_0} t_J} \ket{\psi(0)} +O(V^2)$, the scattering process that absorbs a laser photon and emits a system photon [Eq.~(\ref{p0p})] is the leading-order effect. Therefore, $\sum_i \mathcal{P}_i \approx \mathcal{P}_0(\bp)$, and the normalized probability distribution is thus $P_j \equiv \mathcal{P}_j/\sum_i \mathcal{P}_i\approx \mathcal{P}_j/\mathcal{P}_0(\bp)$. The transition rate associated with $\mathcal{P}_j$
can be found by $\Gamma_j(\bp)= \int_0^1 \gamma(\bp) P_j \mathrm{d}r$, which is the overall decay rate $\gamma(\bp)$ times the normalized probability $P_j$ averaging over possible jump time.

The leading-order jump outcome associated with Eq.~(\ref{p0p}) happens at a rate $\Gamma_0(\bp)=\gamma(\bp)\int_0^1 \mathrm{d}r \frac{\mathcal{P}_0(\bp)}{\mathcal{P}_0(\bp)}=\gamma(\bp)$. This corresponds to the laser-bath scattering process [Fig.\ref{fig:transitions}(a)] with a rate consistent with our analysis in Sec~\ref{sec:twomodecooling}. We can identify this rate as the ``thermalizing jump rate," the jump rate that leads to the thermalization of atoms. In addition, since the laser-bath scattering leads to Doppler cooling of atoms, we can assume that the atomic motion reequilibrates to a steady state $\rho^B_{\rm atom} = \int \mathrm{d}^3 \bp \Pi(\bp)\ket{g,\bp}\bra{g,\bp}$ before other processes involving the change of system photonic state occurs, where $\Pi(\bp)$ follows the Boltzmann distribution as defined in Eq.~(\ref{Boltzmann}). This steady-state distribution due to Doppler cooling of atoms averages out the phase factor $e^{i(\bk_L-\bq)\cdot\br}$ in the coupling $V_{ASL}(t)$; one can thus neglect the coherent part of the system photons. The long-time dynamics of the system photons is then governed by incoherent transitions between photon number states with rates calculated below. 

The average rate of emitting one system photon from a specific atomic momentum state $\ket{g,\bp}$, start from $\ket{g_0}$ and ending in $\ket{g^+_{\bq,L}}$ on top of the rapid jumps [Fig.~\ref{fig:transitions}(b)], is
\begin{align}
&\Gamma^{+}_{\bq, L}(\bp) \equiv \gamma(\bp)\int \frac{\mathcal{P}^+_{\bq ,L}(\bp)}{\mathcal{P}_0(\bp)}  \mathrm{d}r \notag\\
&= \gamma(\bp+\hbar \bk_L-\hbar \bq)|R^+_\bq(\bp)|^2(n_\bq+1)  \int_0^1 \frac{\mathrm{d}r |f_{g^+_{\bq,L}}(r)|^2}{r} .
\label{Gamma+L}
\end{align}
This rate corresponds to the emission process of a system photon over a finite time before the atoms being reset (thermalized) by the emission process into bath modes.
We see a simpler interpretation here: a new jump operator that acts directly on the system photon state, with a jump rate $\kappa^+_{\bq,L}(\bp) = \Pi(\bp) \gamma(\bp+\hbar \bk_L-\hbar \bq)|R^+_\bq(\bp)|^2 \int_0^1 |f_{g^+_{\bq,L}}(r)|^2 \mathrm{d}r/r$ and a jump term $a^{\dag}_{\bq}$.

Evaluating $\int_0^1 |f_{g^+_{\bq,L}}(r)|^2 \mathrm{d}r/r$, we have
\begin{align}
& \int_0^1 \frac{\mathrm{d}r|f_{g^+_{\bq,L}}(r)|^2}{r} =  \gamma(\bp) \int_0^\infty |f_{g^+_{\bq,L}}(t_J)|^2 \mathrm{d}t_J\notag\\
& = \frac{\gamma_{g_0}+ \gamma_{g^+_{\bq,L}}}{\gamma_{g^+_{\bq,L}} \left[ (\omega_{g_0} +\omega_L - \omega_{g^+_{\bq,L}})^2 + \frac{\left(\gamma_{g_0} + \gamma_{g^+_{\bq,L}}\right)^2}{4} \right]},
\label{fgq+L}
\end{align}
where we used $r = e^{-\gamma(\bp) t} \rightarrow \frac{\mathrm{d}r}{r} = - \gamma(\bp) \mathrm{d}t$.
The total system photon emission rate is $(n_{\bq}+1)\Lambda^+_{\bq,L}$ with
\begin{align}
&\Lambda^+_{\bq,L}\equiv \int \mathrm{d}^3 \bp \kappa^+_{\bq,L}(\bp) \notag\\&= \int \mathrm{d}^3 \bp\, \Pi(\bp) |R^+_\bq(\bp)|^2 \delta_{\gamma(\bp) + \gamma(\bp+ \bk_L-\bq)}\left(\Delta E_{gg}(\bq,\bp)\right).
\label{App:lambda_L}
\end{align}
Recall that $\delta_{\epsilon}(\omega)=\frac{\epsilon /2 \pi}{\omega^2 + \epsilon^2/4}$ is a broadened $\delta$ function of $\omega$ with a width $\epsilon$. We call the result in  Eq.~(\ref{App:lambda_L}) an example of the \textit{self-consistent Fermi's golden rule}, in which the $\delta$ function in the usual Fermi's golden rule is now replaced by the decay-broadened $\delta$ function $\delta_{\gamma(\bp) + \gamma(\bp+ \bk_L-\bq)}$ due to the finite lifetime of the initial and final states. The system photons absorption rates according to Eqs.~(\ref{p-Lp}) and (\ref{p-Bp}) can be found analogously as presented in Sec.~\ref{sec:SCFGR}.

\bibliographystyle{apsrev4-1}
\bibliography{thermal}

\end{document}